\documentclass[usenatbib]{mn2e}

\usepackage{graphicx}
\usepackage{amssymb}

\title[Are the missing X-ray breaks in GRB afterglow light curves merely hidden?]
{Are the missing X-ray breaks in Gamma-ray Burst afterglow light curves merely hidden?}

\author[P.A.~Curran et al.]
{P.A.~Curran$^1$\thanks{e-mail: pcurran@science.uva.nl}, 
A.J.~van~der~Horst$^{1,2}$,
R.A.M.J.~Wijers $^1$\\
$^1$Astronomical Institute, University of Amsterdam, Kruislaan 403, 1098\,SJ Amsterdam, The Netherlands \\
$^2$ University of Alabama, National Space Science and Technology Center, 320 Sparkman Drive, Huntsville, AL 35805, USA\\
}

\begin{document}

\date{Accepted 2008 January 29. Received 2008 January 14; in original form 2007 October 26}

\pagerange{\pageref{firstpage}--\pageref{lastpage}} \pubyear{}

\maketitle

\label{firstpage}


\begin{abstract}
Gamma-ray burst afterglow observations in the \emph{Swift} era have a perceived lack of achromatic jet breaks compared to the \emph{BeppoSAX} or pre-\emph{Swift} era. Specifically, relatively few breaks, consistent with jet breaks, are observed in the X-ray light curves of these bursts. If these breaks are truly missing, it has serious consequences for the interpretation of GRB jet collimation and energy requirements, and the use of GRBs as cosmological tools.
Here we address the issue of X-ray breaks that are possibly `hidden' and hence the light curves are misinterpreted as being single power laws. We do so by synthesising XRT light curves and fitting both single and broken power laws, and comparing the relative goodness of each fit via Monte Carlo analysis. Even with the well sampled light curves of the \emph{Swift} era, these breaks may be left misidentified, hence caution is required when making definite statements on the absence of achromatic breaks. 
\end{abstract}



\section{Introduction}\label{section:introduction}

The afterglow emission of Gamma-Ray Bursts (GRBs) is well described by the blast wave, or fireball, model \citep{rees1992:mnras258,meszaros1998:apj499}. This model details the temporal and spectral behaviour of the emission that is created by external shocks when a collimated ultra-relativistic jet ploughs into the circumburst medium, driving a blast wave ahead of it. 
The level of collimation, or jet opening angle, has important implications for the energetics of the underlying physical process and the possible use of GRBs as standard candles. 
The signature of the collimation, according to simple analytical models, is an achromatic temporal steepening  or `jet break' at $\sim 1$ day in an otherwise decaying, power law light curve; from the time of this break, the jet opening angle can be estimated \citep{rhoads1997:ApJ487}.

Since the launch of the \emph{Swift} satellite \citep{gehrels2004:ApJ611}, this standard picture of afterglows has been called into question by the lack of observed achromatic temporal breaks, up to weeks in a few bursts (e.g., \citealt{panaitescu2006:MNRAS369,burrows2007:astro.ph2633}). In some afterglows, a break is unobserved in both the X-ray and optical light curves, while in other bursts a break is observed in one regime but not the other (e.g., \citealt{liang2007:arXiv0708}).
In the \emph{BeppoSAX} era, most well sampled light curves were in the optical regime, while in the \emph{Swift} era most well sampled light curves are in the X-ray regime. Our expectations of the observable signature of a jet break are hence based on the breaks observed pre-\emph{Swift}, predominately by optical telescopes, and the models which explained them, notably those of \cite{rhoads1997:ApJ487,rhoads1999:ApJ525} and \cite{sari1999:ApJ519}. 
It is not clear that the breaks will be identical in both the X-ray and optical regimes. 
In the cases of GRB\,990510 and GRB\,060206, both have clear breaks in the optical but only marginal breaks in the X-ray (\citealt{kuulkers2000:ApJ538,curran2007:MNRAS381} respectively). Regardless of this, GRB\,990510 is taken as a prototypical achromatic jet break, while the achromatic nature of the GRB\,060206 break is only evident  when supported by the broad-band spectral indices.

In this paper we address the issue of X-ray breaks which are possibly `hidden' and hence the light curve misinterpreted as being a single power law. We do so by synthesising X-ray light curves and fitting both single and broken power laws, and comparing goodness of each fit via the F-test. 
In \S\ref{section:method} we introduce our method while in \S\ref{section:results} we present the results of our Monte Carlo analysis. In \S\ref{section:discussion} we discuss the implications of these results in the overall context of GRB observations and we summarise our findings in \S\ref{section:conclusion}.


\section{Method}\label{section:method}

\subsection{Models}\label{section:models}

In accordance with the fireball model, we ascribe the behaviour of the X-ray light curve to be a single power law decay where the flux goes as:
\[
F_{\nu}(t) \propto  t^{-\alpha}, 
\]
with a temporal decay index, $\alpha$. Alternatively, if there is a break we use a smoothly broken power law decay with a break at $t_b$: 
\[
\label{eq:Beuermann}
F_\nu(t) = F_\nu(t_b) 
\left[\left(\frac{t}{t_b}\right)^{\alpha_1 S}+\left(\frac{t}{t_b}\right)^{\alpha_2 S}\right]^{-1/S}, 
\]
where $S$ is the smoothing factor (a larger value corresponding to a sharper break) 
and the light curve goes as $F_{\nu} \propto t^{-\alpha_1}$ and $F_{\nu} \propto t^{-\alpha_2}$ pre- and post-break respectively ($\alpha_1 < \alpha_2$).

We have simulated  both single and broken power law data  for a limited, but varied number of realistic parameter sets. 
The flux offsets are set at a reasonably high rate of 0.1\,cts/s at 1\,day to give well sampled light curves. 
We take break times at 0.35, 1 and 2\,days, and input smoothing parameters, $S = 1, 2, 5$. 
The temporal decay indices are chosen as a function of electron energy distribution index, $p$, assuming i) X-ray frequency, $\nu_{\mathrm{X}} > \nu_{\mathrm{c,m}}$, ii)  $\nu_{\mathrm{m}} < \nu_{\mathrm{X}} < \nu_{\mathrm{c}}$ in a wind-like environment  and iii) $\nu_{\mathrm{m}} < \nu_{\mathrm{X}} < \nu_{\mathrm{c}}$ in an ISM-like (constant density) environment, where $\nu_{\mathrm{c}}$ and $\nu_{\mathrm{m}}$ are the synchrotron cooling and peak frequencies (e.g. \citealt{zhang2004:IJMPA19}). 
We adopt three values of $p =1.8, 2.4, 3.0$ covering the observed range of $p$ values (e.g. \citealt{panaitescu2001:ApJ560,starling2007:arXiv0704.3718}), which imply the temporal indices, $\alpha$ as shown in Table\,\ref{alphas}.

\begin{table}	
  \centering	
  \caption{The values of temporal indices, $\alpha$, for the given values of $p$ in the given cases, from  \protect\cite{zhang2004:IJMPA19}. } 	
    \label{alphas} 	
    \begin{tabular}{l l l l} 
      \hline 
      $p$        &  1.8  & 2.4  &  3.0 \\
      \hline \hline
      $\nu_{\mathrm{X}} > \nu_{\mathrm{c,m}}$                       & 0.96   & 1.30   &  1.75      \\
      $\nu_{\mathrm{m}} < \nu_{\mathrm{X}} < \nu_{\mathrm{c}}$ (Wind) & 1.23   & 1.55   &  2.00      \\
      $\nu_{\mathrm{m}} < \nu_{\mathrm{X}} < \nu_{\mathrm{c}}$ (ISM)  & 0.71   & 1.05   &  1.50       \\    
      \hline
      post-break                & 1.95   & 2.40   &  3.00       \\    
      \hline  \hline 
    \end{tabular}
\end{table}

\subsection{Synthesis of data}\label{section:synth}

In synthesising \emph{Swift} X-Ray Telescope (XRT; \citealt{burrows2005:SSRv120}) light curves we wanted them to be comparable to those produced by the online repository \citep{evans2007:A&A469}, so we tried to satisfy the criteria that were applied to their production process. It is therefore instructive to describe briefly their method, and the differences to ours. 
Firstly, \citeauthor{evans2007:A&A469} present a reduction method dealing with observed source and background counts whereas ours deals directly with the rate predicted by an underlying model. \citeauthor{evans2007:A&A469} demand that each bin (i.e., data point) must have a minimum time span and number of counts from the source region. The minimum number of counts per bin is dependent on which mode the XRT is observing in (i.e., Windowed Timing/Photon Counting (WT/PC); \citealt{hill2004:SPIE5165,hill2005:SPIE.5898}) and the measured counts of the source region. 
As we will be dealing with late time data ($> 10^4$\,s) which is predominately measured in the PC mode, and has low count rates ($<1$\,cts/s), we use a constant value for the counts per bin. We do this by setting the bin interval so that the approximate numerical integration of the rate, i.e., the counts, is constant.
Also at these late times, it is unlikely that this minimum count will be reached in less than the minimum time span so the time span criterion may be ignored.

The main conditions of our synthesised light curves are as follows: 
\begin{itemize}
\item{Constant counts, and $1\sigma$ fractional error of 0.25, per data point}
\item{94\,minute orbits (47\,min on/off)}
\item{Fractional exposure of 0.1 after 1 day}
\item{Rate cut off at $5 \times 10^{-4}$\,cts/s}
\item{Time spans $10^4$ to $2.6 \times10^6$\,s (3\,hours to 30\,days)}
\end{itemize}

Our constant counts per data point is not comparable to the constant used by \citeauthor{evans2007:A&A469}, though we have chosen a value so that the number of data points over our light curve is comparable to XRT light curves from the repository.  The error value of an individual data point is taken as a constant (since the counts per bin are constant) fractional value of 0.25, to agree with the values obtained from a number of well sampled XRT light curves in \citeauthor{evans2007:A&A469}. 

As the \emph{Swift} satellite is in a low Earth orbit with a period of 94\,minutes it suffers from 50 per cent time off target per orbit. We take this into account by not synthesizing data for these off-target periods, assuming that we start our light curve at the start of the on-target time. Even after this orbital time loss, \emph{Swift} does not dedicate 100 per cent of the time that the target is visible to the target itself. This is due to other observing commitments and constraints. In fact the fractional exposure or actual fraction of time on target after the orbital constraint, is 0.1 after $\sim 1$\,day for most bursts in \citeauthor{evans2007:A&A469}, leading to $\sim 5$\,minutes of exposure time per orbit.

As we are only interested in the power-law decay or afterglow phase of GRBs, we start our synthesis at $1 \times 10^4$\,s. This is a time after which many real XRT light curves have data unaffected by early flaring or prompt emission \citep{chincarini2007:ApJ671}. 
XRT can sometimes continue to observe for up to a month ($\sim 2.6 \times 10^6$\,s) so we synthesise data out to that same time, unless it drops below a threshold of $5 \times 10^{-4}$\,cts/s before then. 

\begin{figure} 
  \centering 
  \resizebox{\hsize}{!}{\includegraphics[angle=-90]{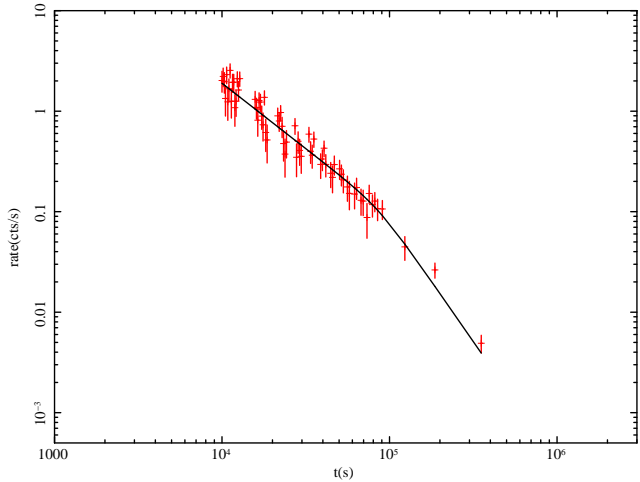} }
  \caption{A sample synthesised XRT (SXRT) data set including associated errors, overlaid on the underlying smoothly broken power law with a break at 1\,day, smoothing factor, $S=5$ and $p=2.4$ ($\nu_{\mathrm{X}} > \nu_{\mathrm{c,m}}$).} 
  \label{sample-broken} 
\end{figure}

\begin{figure} 
  \centering 
  \resizebox{\hsize}{!}{\includegraphics[angle=-90]{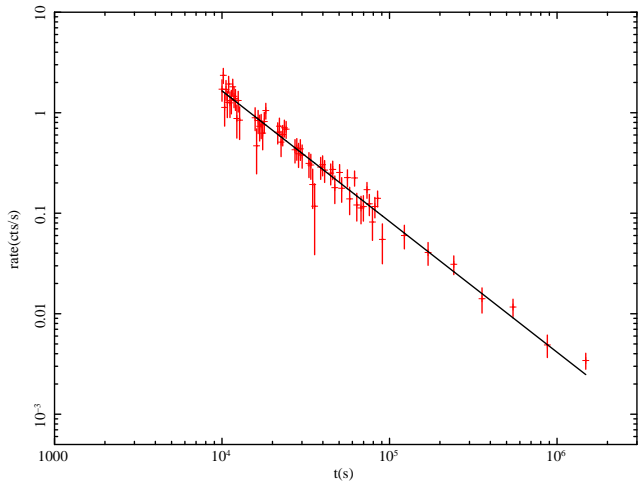} }
  \caption{A sample synthesised XRT (SXRT) data set including associated errors, overlaid on the underlying single power law with  $p=2.4$ ($\nu_{\mathrm{X}} > \nu_{\mathrm{c,m}}$).} 
  \label{sample-single} 
\end{figure}

From the data points that lie on the model curves, the simulated data, we synthesise the final data points by randomly perturbing the simulated data within the Gaussian errors. We simulated both single and broken power law light curves from all possible combinations of the parameters defined in section \ref{section:models}. Samples of two synthesised XRT (SXRT) light curves are shown in Figures \ref{sample-broken} \& \ref{sample-single}.

\subsection{Monte Carlo analysis: Fitting of data and the F-test}\label{section:fitting}

The SXRT data, defined  by given set of input parameters, are fit automatically using the \emph{simulated annealing} method (\S 10.9 of \citealt{press1992:nrca} and references therein) to minimise the $\chi^2$ of the residuals. The SXRT data are first fit to a single power law, then a smoothly broken power law of fixed smoothing parameter, $s$, and the two are compared via an F-test to evaluate their relative suitability. 
The data points are then re-perturbed from their original, on-model, simulated values and refit repeatedly (1000 trials) in a Monte Carlo analysis to get average values and $1\sigma$ Gaussian deviations of the best fit parameters and F-test probabilities. This process is then repeated for all possible single and broken power laws allowed by the possible combinations of the parameters in section \ref{section:models}.


\section{Results}\label{section:results}

\subsection{The F-test}\label{section:results:f-test}

Before we discuss our results, let us first consider the interpretation of the F-test, which returns an F-test statistic that we convert to an F-test probability, $\rm{F_{prob}}$. In this situation, the F-test is a measure of the probability that the decrease in $\chi^2$ associated with the addition of the two extra parameters of a broken power law, $\alpha_2$ and $t_{\mathrm{break}}$, is by chance or not. As is common practice in light curve analysis literature, the following, though arbitrary, are approximately true:
\begin{itemize}
\item{$\rm{F_{prob}} \gtrsim 10^{-2} \Rightarrow$ favours single power law}
\item{$10^{-5} \lesssim \rm{F_{prob}} \lesssim 10^{-2} \Rightarrow$ neither favoured}
\item{$\rm{F_{prob}} \lesssim 10^{-5}  \Rightarrow$ favours broken power law}
\end{itemize}

We presume a broken power law only if the probability of a chance improvement of the $\chi^2$ is very low ($\rm{F_{prob}} \lesssim 10^{-5}$), while a single power law is presumed if the probability of a chance improvement is very high ($\rm{F_{prob}} \gtrsim 10^{-2}$). For the intermediate cases with moderate probability ($10^{-5} \lesssim \rm{F_{prob}} \lesssim 10^{-2}$) of a chance improvement, a single power law is usually presumed as it is the simpler model. An $\rm{F_{prob}} = 10^{-2}$ corresponds to a $3\sigma$ detection of a break while $\rm{F_{prob}} = 10^{-5}$ corresponds to slightly less than $5\sigma$ certainty.
We have modified the F-test somewhat to return probabilities greater than 1 when the $\chi^2$ of the broken power law is greater than the single power law. If the underlying power law is single, a broken  power law may give a worse fit as it is defined to break to a steeper slope only ($\alpha_1 < \alpha_2$), not a shallower slope as may occur during the data synthesis.

\begin{figure} 
  \centering 
  \resizebox{\hsize}{!}{\includegraphics[angle=0]{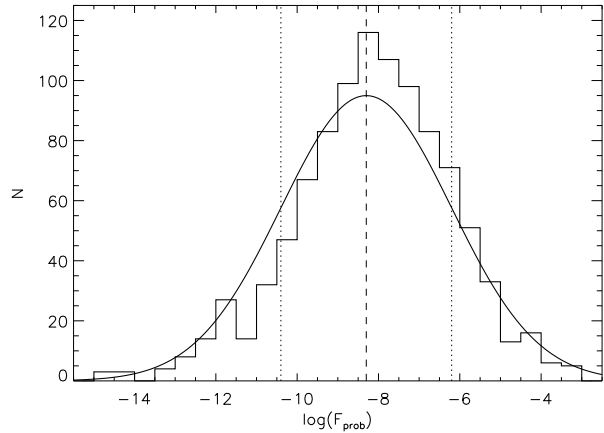} }
  \caption{The distribution of the log of F-test probabilities, $\log(\rm{F_{prob}})$ for the SXRT data set described in the caption of Figure \ref{sample-broken} ($p=2.4$, $\nu_{\mathrm{X}} > \nu_{\mathrm{c,m}}$, $t_{\mathrm{break}} =1$\,day, smoothing, $S=5$). The vertical lines show the average value and $1\sigma$ distribution, $\log(\rm{F_{prob}}) = -8.3 \pm 2.1$ (as in Table \ref{broken_log}) of the data while the curve is the Gaussian with these values, for comparison.}
  \label{sample-distribution} 
\end{figure}

When fitting the data, we use a number of different smoothing factors ($s = 1, 2, 5, 10$) and find that the F-test probability does not have a strong dependency, if any, on the smoothing factor of the model used to fit the data, only on the smoothing factor used to generate the data, $S$. For this reason we present only our F-test results for fits with a smoothing factor $s = 5$. 
The values of $\log(\rm{F_{prob}})$ given in Tables \ref{broken_log}, \ref{broken_log_wind} \& \ref{broken_log_ism} show that a true, underlying broken power law could be mistaken ($-5 \lesssim \log(\rm{F_{prob}}) \lesssim -2$, bold font) for a single power law in a number of cases. Due to the $1\sigma$ distribution of the returned F-test probabilities for a given set of parameters, it is clear that even in cases with $\rm{F_{prob}} \lesssim 10^{-5}$, a proportion could be misidentified as single power laws (Figure \ref{sample-distribution}).

For a given value of $p$, those data sets where the underlying power law has a quite smooth transition from the asymptotic values ($S \sim 1$), are most likely to be mistaken for single power laws. 
If the break is at the edge of the time range, it is also more likely to be misidentified and this is especially true in the case of a break at later times, when there are few data points post-break. 
There are a  number of conflicting dependencies on the value of $p$ itself. 
Firstly, the difference between the two temporal decay indices, $\Delta\alpha$ ($p > 2$): 
\begin{itemize}
\item{$\Delta\alpha =  (2+p)/4$ ($\nu_{\mathrm{X}} > \nu_{\mathrm{c,m}}$)}
\item{$\Delta\alpha =  (1+p)/4$ ($\nu_{\mathrm{m}} < \nu_{\mathrm{X}} < \nu_{\mathrm{c}}$, Wind)}
\item{$\Delta\alpha =  (3+p)/4$ ($\nu_{\mathrm{m}} < \nu_{\mathrm{X}} < \nu_{\mathrm{c}}$, ISM)}
\end{itemize}
so that a larger value of $p$ implies a more pronounced, easily detectable break. However a larger value of $p$ also causes steeper decay indices and fewer SXRT data points post break, making the break more difficult to confirm. If the break occurs at an early enough time (e.g., $t_{\mathrm{break}}$ = 0.35\,days) so that there is enough data post-break, this effect is unlikely to be an issue.
The values of $\Delta\alpha$ are also dependent on the regime and environment, and we also observe a strong dependency on these. It is clear from the results for a constant density medium with $\nu_{\mathrm{m}} < \nu_{\mathrm{X}} < \nu_{\mathrm{c}}$ (Table \ref{broken_log_ism}) that there are far fewer misidentifications than in the results for a wind-like medium with $\nu_{\mathrm{m}} < \nu_{\mathrm{X}} < \nu_{\mathrm{c}}$ (Table \ref{broken_log_wind}).
This may introduce a bias in the detected breaks of GRBs as those in a constant density medium ($\nu_{\mathrm{m}} < \nu_{\mathrm{X}} < \nu_{\mathrm{c}}$) are most obvious and hence easier to identify.

Looking at the results for the single power laws in Tables \ref{broken_log}, \ref{broken_log_wind} \& \ref{broken_log_ism} it is clear that if the underlying power law is not broken then $\log(\rm{F_{prob}}) \sim 0$ or $\rm{F_{prob}} \sim 1$ and it is highly unlikely to be mistaken for a broken power law, even considering the distribution of the results. Given that the cases where intermediate probabilities ($-5 \lesssim \log(\rm{F_{prob}}) \lesssim -2$) are found all have underlying broken power laws, if a broken power law does offer an improved fit then the underlying power law is likely broken.

The above however, obviously depends on the intensity of the burst and dimmer bursts will have less certain conclusions, while the brighter bursts are less likely to have misidentifications. By normalising all our SXRT light curves at 1\,day, we have made them as comparable as possible. 
We have simulated a number of representative bursts with physical parameters, which show that in the circumstances described above, a broken power law may be `hidden', i.e., an F-test will not favour the broken power law over the single power law, though the underlying power law is indeed broken.


\begin{table}	
  \centering	
  \caption{Log of F-test probabilities, $\log(\rm{F_{prob}})$ for SXRT data with given values of $p$ and $\nu_{\mathrm{X}} > \nu_{\mathrm{c,m}}$. The SXRT data are generated from a single power law, and from smoothly broken power laws with given values of $S$, and breaks at 0.35, 1 and 2\,days. 
The cases where the F-test favours neither fit model ($-5 \lesssim \log(\rm{F_{prob}}) \lesssim -2$), so a single power law would normally be assumed, are in bold font.}
  \label{broken_log} 	
  \begin{tabular}{l c c c} 
    $t_{\mathrm{break}}$ & & & \\
    & $p = 1.8$          & $p = 2.4$            & $p = 3.0$ \\ 
    \hline \hline
    no $t_{\mathrm{break}}$ &&&\\
    & -0.10 $\pm$ 0.42 & -0.10 $\pm$ 0.31  &  -0.07 $\pm$ 0.34   \\    
    \hline 
    0.35\,days &&&\\
    $S=1$     & -5.3 $\pm$ 1.8  & {\bf-4.4 $\pm$ 1.8}  & -10.0 $\pm$ 3.2  \\
    $S=2$     & -7.0 $\pm$ 2.1  & -9.6 $\pm$ 2.6  & -14.7 $\pm$ 4.3 \\    
    $S=5$     &-11.1 $\pm$ 2.4  & -16.5 $\pm$ 3.2 & -27.1 $\pm$ 6.3  \\    
    \hline 
    1\,day &&&\\ 
    $S=1$      & {\bf-4.5 $\pm$ 1.6}  & -6.7  $\pm$ 1.9  & -7.6 $\pm$ 2.3  \\    
    $S=2$      & -8.4 $\pm$ 1.0  & -11.7 $\pm$ 2.4  & -13.2 $\pm$ 2.8 \\    
    $S=5$      & -7.9 $\pm$ 2.0  &  -8.3 $\pm$ 2.1  & -16.4 $\pm$ 2.8  \\    
    \hline 
    2\,days &&&\\
    $S=1$      & {\bf-4.9 $\pm$ 1.6}    & {\bf-3.9 $\pm$ 1.7}  & {\bf-3.8 $\pm$ 1.7}  \\    
    $S=2$      & -7.1 $\pm$ 1.8   & {\bf-4.8 $\pm$ 1.7}        & {\bf-3.7 $\pm$ 1.6}  \\    
    $S=5$      & -7.3 $\pm$ 1.9   & {\bf-4.3 $\pm$ 1.6}        & {\bf-2.7 $\pm$ 1.5}  \\    
    \hline \hline 
  \end{tabular}
\end{table}

\begin{table}	
  \centering	
  \caption{Same as Table \ref{broken_log} but for the case that $\nu_{\mathrm{m}} < \nu_{\mathrm{X}} < \nu_{\mathrm{c}}$ in a wind-like environment. }
  \label{broken_log_wind} 	
  \begin{tabular}{l c c c} 
    $t_{\mathrm{break}}$ & & & \\
    & $p = 1.8$          & $p = 2.4$            & $p = 3.0$ \\ 
    \hline \hline
   no $t_{\mathrm{break}}$ &&&\\
  & -0.13 $\pm$ 0.36 & -0.09 $\pm$ 0.38  &  -0.08 $\pm$ 0.37   \\    
    \hline 
    0.35\,days &&&\\
    $S=1$     & {\bf-2.4 $\pm$ 1.4} &  {\bf-2.0 $\pm$ 1.3}  &  -5.8 $\pm$ 2.4   \\
    $S=2$     & {\bf-3.4 $\pm$ 1.6} &  -5.2 $\pm$ 2.0  &  -7.7 $\pm$ 3.1   \\    
    $S=5$     & -6.5 $\pm$ 2.1 & -10.4 $\pm$ 2.7  & -17.4 $\pm$ 5.6   \\    
    \hline
    1\,day &&&\\
    $S=1$      & {\bf-2.3 $\pm$ 1.3}   & {\bf-3.9 $\pm$ 1.8}  &  {\bf-4.3 $\pm$ 2.0}     \\    
    $S=2$      & -5.4 $\pm$ 1.9   & {\bf-3.6 $\pm$ 1.7}  &  -9.2 $\pm$ 2.8     \\    
    $S=5$      & -5.4 $\pm$ 1.8   & -5.8 $\pm$ 2.0  & -13.4 $\pm$ 3.1     \\    
    \hline 
    2\,days &&&\\
    $S=1$      & {\bf-1.8 $\pm$ 1.2}  & {\bf-2.5 $\pm$ 1.4}  &  {\bf-2.4 $\pm$ 1.6}  \\    
    $S=2$      & {\bf-3.1 $\pm$ 1.5}  & {\bf-3.9 $\pm$ 1.7}  &  {\bf-3.4 $\pm$ 1.6}  \\    
    $S=5$      & {\bf-3.8 $\pm$ 1.6}  & {\bf-4.2 $\pm$ 1.8}  &  {\bf-2.6 $\pm$ 1.5}  \\    
    \hline \hline 
  \end{tabular}
\end{table}

\begin{table}	
  \centering	
  \caption{Same as Table \ref{broken_log} but for the case that $\nu_{\mathrm{m}} < \nu_{\mathrm{X}} < \nu_{\mathrm{c}}$ in an ISM-like (constant density) environment. }
  \label{broken_log_ism} 	
  \begin{tabular}{l c c c} 
    $t_{\mathrm{break}}$ & & & \\
    & $p = 1.8$          & $p = 2.4$            & $p = 3.0$ \\ 
    \hline \hline
   no $t_{\mathrm{break}}$ &&&\\
  & -0.05 $\pm$ 0.56 &  -0.08 $\pm$ 0.39  &   -0.09 $\pm$ 0.34  \\    
    \hline
    0.35\,days &&&\\
    $S=1$     &  -8.8 $\pm$ 2.1  &  -7.8 $\pm$ 2.2  &  -19.6 $\pm$ 3.7    \\
    $S=2$     & -10.5 $\pm$ 2.2  & -14.5 $\pm$ 2.8  &  -22.4 $\pm$ 5.1  \\    
    $S=5$     & -15.1 $\pm$ 2.6  & -21.4 $\pm$ 3.2  &  -35.8 $\pm$ 6.7   \\    
    \hline
    1\,day &&&\\
    $S=1$      &  -6.7 $\pm$ 1.8  &  -9.3 $\pm$ 2.1  & -11.9 $\pm$ 2.5  \\    
    $S=2$      & -10.2 $\pm$ 1.9  & -13.9 $\pm$ 2.3  & -17.2 $\pm$ 2.7  \\    
    $S=5$      &  -9.2 $\pm$ 1.9  &  -9.8 $\pm$ 2.1  &  -7.8 $\pm$ 2.1  \\    
    \hline
    2\,days &&&\\
    $S=1$      &  -8.4 $\pm$ 1.7   &  -8.9 $\pm$ 2.0  &  {\bf-4.5 $\pm$ 1.9} \\    
    $S=2$      & -10.1 $\pm$ 1.9   & -10.1 $\pm$ 2.0  &  {\bf-3.7 $\pm$ 1.7}  \\    
    $S=5$      & -10.4 $\pm$ 1.8   &  -9.3 $\pm$ 1.9  & -11.9 $\pm$ 2.3 \\    
    \hline \hline 
  \end{tabular}
\end{table}


\subsection{Return of parameters}\label{section:results:return}

Unlike in the case of the F-test probabilities, which have little or no dependency on the smoothing factor, $s$, used to fit the data, the returned values of temporal slopes, $\alpha_1$ \& $\alpha_2$ and break time, $t_{\rm{break}}$ do show some, though only marginally significant, dependency on this. 
To  demonstrate this effect we show the case which gives a median value of $\Delta\alpha$ ($\nu_{\mathrm{X}} > \nu_{\mathrm{c,m}}$, $p = 2.4$) and use a standard break time of 1\,day (Table \ref{return_parameters}), though this discussion also holds for the other break times in our synthesis.

We see from the best fit parameters returned from a given SXRT data set, that data fit with smoother breaks ($s=1$) give later $t_{\rm{break}}$, shallower $\alpha_1$ and steeper  $\alpha_2$ than those fit with sharper breaks ($s=5$). Fitting the data with a smoothing factor, $s=S$ seems to give the most accurate values of the returned parameters, though in the case that the break is smooth ($S=1$), the returned values of especially $\alpha_2$ and $t_{\rm{break}}$ are not very accurate. As an example, we can see from the case of $S=5$ and $s=1$ that fitting with a smoothing parameter less than the real smoothness ($s<S$) will underestimate $\alpha_1$ while overestimating $\alpha_2$ and $t_{\rm{break}}$. The reverse is also true, in that if the data are fit with a smoothing parameter greater than the real smoothness ($s>S$, e.g., $S=1$ and $s=5$), $\alpha_1$ is overestimated while $\alpha_2$ and $t_{\rm{break}}$ are underestimated somewhat.

Whether or not $s=S$, there is quite a spread in the returned parameters for a given set of input parameters. This is similar to the result of \cite{johannesson2006:ApJ640}, which suggests caution in the interpretation of results of broken power law fits to light curves. It could be an influence on the observed deviation of post-break slopes from $\alpha_2 = p$.

\begin{table}	
  \centering	
  \caption{The best fit parameters for a set of SXRT data for various values of fit smoothing parameter, $s$. The SXRT data is generated from a broken power law with $p = 2.4$ ($\nu_{\mathrm{X}} > \nu_{\mathrm{c,m}}$), a break at 1\,day, and given values of smoothing parameter, $S$.} 	
    \label{return_parameters} 	
    \begin{tabular}{l l l l} 
                  & $\alpha_{1}$ & $\alpha_{2}$      & $t_{\rm{break}} \times 10^4$\,s  \\ 
      \hline 
      input       &  1.30  & 2.4  &  8.64 \\
      \hline \hline
      $S=1$&&&\\  
      $s=1$      & 1.30 $\pm$ 0.18   & 2.53 $\pm$ 0.37   &  12.5 $\pm$ 8.3     \\
      $s=2$      & 1.38 $\pm$ 0.15   & 2.32 $\pm$ 0.26   &   9.5 $\pm$ 6.1      \\
      $s=5$      & 1.45 $\pm$ 0.11   & 2.20 $\pm$ 0.28   &   8.1 $\pm$ 5.4      \\    
      \hline 	     
      $S=2$&&&\\     
      $s=1$      & 1.19 $\pm$ 0.16   &  2.66 $\pm$ 0.28  &  11.6 $\pm$ 5.9   \\
      $s=2$      & 1.28 $\pm$ 0.11   &  2.44 $\pm$ 0.20  &  9.3 $\pm$ 3.7    \\
      $s=5$      & 1.35 $\pm$ 0.08   &  2.32 $\pm$ 0.20  &  8.3 $\pm$ 3.2   \\    
      \hline 	     
      $S=5$&&&\\     
      $s=1$      & 1.20 $\pm$ 0.12   & 3.27 $\pm$ 0.63   &  14.6 $\pm$ 4.7   \\
      $s=2$      & 1.25 $\pm$ 0.09   & 2.78 $\pm$ 0.40   &  11.0 $\pm$ 3.1    \\
      $s=5$      & 1.29 $\pm$ 0.07   & 2.50 $\pm$ 0.35   &  9.0 $\pm$ 2.5   \\    
      \hline  \hline 
    \end{tabular}
\end{table}



\section{Discussion}\label{section:discussion}

We have shown that even an ideal, non-flaring, well sampled X-ray light curve synthesised from an underlying broken power law can, in some cases, be as satisfactorily fit by a single power-law decay as a broken power-law. 
This would lead us, in the absence of further evidence, to assume a single power law and rule out the possibility of a break. 
This effect is most pronounced in bursts with late break times or high levels of smoothing ($S \sim 1$), as many previously studied jet breaks display (e.g., \citealt{zeh2006:ApJ637}). In the cases of high levels of smoothing, the breaks do not display sharp changes in the temporal decay index but a shallow roll-over from asymptotic values, well described by a smoothly broken power-law, which makes accurate determination of the temporal slopes and break time difficult.

The prototypical example of such a break is GRB\,990510 for which well sampled $B$, $V$, $R$ and $I$ band light curves display an achromatic break (e.g., \citealt{stanek1999:ApJ522}). 
This is accepted as a jet break even though the X-ray light curve as measured by \emph{BeppoSAX} \citep{kuulkers2000:ApJ538} is satisfactorily described by a single power-law, though the possibility of a break is not eliminated. Likewise, in the case of GRB\,060206 \citep{curran2007:MNRAS381}, the X-ray break is not well pronounced but supported by spectral and optical data.

This has significant implications for the analysis of the numerous X-ray light curves that the \emph{Swift} satellite has afforded us. For those X-ray light curves extending up to $\sim 1$\,day or longer, for which we do not have well sampled optical light curves, caution is required when making claims about the absence of breaks based solely on a comparison of the nominal fitted slopes. Multi-wavelength (optical \& X-ray) temporal and spectral data up to late times are required to make a confident statement on the absence or presence of an achromatic break, and to determine the break parameters.
This is particularly important when performing statistical analyses on a large sample of bursts, for making collimation corrected energy estimates, and for using GRBs as standard candles.

Judging from the cases of GRB\,990510 and GRB\,060206, there may be a tendency, if not yet strongly significant, for X-ray light curves to have less pronounced or smoother breaks than optical light curves and more detailed theoretical models of jet breaks are necessary to clarify whether jet breaks could vary somewhat between wavebands.  
A full population synthesis, which is beyond the scope of this paper, is necessary to estimate the percentage of observed bursts that may have hidden X-ray breaks.


\section{Conclusion}\label{section:conclusion}

As underlying smoothly broken power laws may be \emph{hidden} as single power laws, in XRT light curves, we should exercise caution in ruling out breaks based solely on a comparison of the nominal fitted slopes. We hence need to be cautious in implying chromatic breaks from optical and X-ray light curves where an X-ray break is not ruled out with a high degree of certainty. Multi-wavelength temporal and spectral data are required to make a confident statement on the absence or presence of an achromatic break.
There may be a bias towards detecting breaks from bursts with a constant density circumburst medium, as these are most obvious and easily detectable. 

We have shown that the fitted temporal slopes of smoothly broken power laws show significant variation about central values and that, especially in the case of a smooth break, the underlying values are difficult to extract via a fit. More accurate and wavelength specific descriptions of breaks, likely involving numerical simulations of the jet dynamics, are necessary to better understand the observable signature of breaks and to clarify whether the breaks could vary somewhat between wavebands.


\section*{Acknowledgements}
We thank A.P. Beardmore \& K.L. Page for useful discussions on the XRT.
We thank the referee for constructive comments.  
PAC, RAMJW gratefully acknowledge support of NWO under grant 639.043.302.


\label{lastpage}

\end{document}